\documentclass[conference]{IEEEtran}
\IEEEoverridecommandlockouts
\usepackage{multirow}
\usepackage{subfigure}
\usepackage{float}
\usepackage[pdftex]{graphicx}
\usepackage{cite}
\usepackage{amsmath,amssymb,amsfonts}
\usepackage{algorithmic}
\usepackage{graphicx}
\usepackage{textcomp}
\usepackage{xcolor}

\def\BibTeX{{\rm B\kern-.05em{\sc i\kern-.025em b}\kern-.08em
    T\kern-.1667em\lower.7ex\hbox{E}\kern-.125emX}}

\usepackage{graphics} 
\usepackage{epsfig} 
\usepackage{mathptmx} 
\usepackage{times} 
\usepackage{amsmath} 
\usepackage{amssymb}  
\usepackage{multirow}
\usepackage{subfigure}
\usepackage{float}
\usepackage[pdftex]{graphicx}
\usepackage{subfigure}
\usepackage{float}
\usepackage{graphicx}
\usepackage{multirow}
\graphicspath{ {images/} }
\begin{document}
\title{
Wafer Quality Inspection using Memristive  LSTM, ANN, DNN and HTM}



\author{\IEEEauthorblockN{ Kazybek Adam, Kamilya Smagulova, Olga Krestinskaya, Alex Pappachen James}
\IEEEauthorblockA{\textit{Department of Electrical and Computer Engineering} \\
\textit{Nazarbayev University}\\
Astana, Kazakhstan\\
Email: \{apj\}@ieee.org}}


\maketitle

\begin{abstract}

The automated wafer inspection and quality control is complex and time consuming task, which can be speed up using neuromorphic memristive architectures, as a separate inspection device or integrating directly into sensors.
This paper presents the performance analysis and comparison of  different neuromorphic architectures for patterned wafer quality inspection and classification.
The application of non-volatile memristive devices in these architectures ensures low power consumption, small on-chip area scalability. We demonstrate that Long-Short Term Memory (LSTM) outperforms other architectures for the same number of training iterations, and has relatively low on-chip area and power consumption.  


 
\end{abstract}

\section{INTRODUCTION}

With the increase of density and complexity of semiconductor devices on the wafer, wafer surface inspection becomes increasingly complex, important, and time consuming task.
There are various techniques applied to detect wafer defects, including image processing \cite{yoda1988automatic}, optical methods \cite{fairley2007high} and electron beam inspection \cite{sun2010semiconductor}. The automated wafer inspection process can be speed up using the machine learning architectures that can be used as a separate inspection device or integrated directly into the sensing devices, which can analyze the wafer without manual inspection and sending data to the computer for software processing.  
\par
In this paper, we investigate and compare the application of neuromorphic architectures for wafer quality inspection. We present the performance analysis and comparison of wafer classification accuracy, on-chip area and power consumption of a single Perceptron \cite{rosenblatt1958perceptron}, a three-layer Artificial Neural Network (ANN) \cite{8351344}, Long short-term memory (LSTM) neural netwrok \cite{smagulova2018memristor}, Deep Neural Network (DNN) \cite{8060399, 7966300} and Hierarchical Temporal Memory (HTM) \cite{8023844}. The performance of neuromorphic architectures is tested using the database of wafer parameters from \cite{olszewski2001generalized}, consisting of two classes of time-series data obtained during measurement of inline semiconductor processing. 

\section{Neuromorphic architectures for wafer quality inspection}

\begin{figure}[!ht]
\centering
	\subfigure[]{
	\centering
    \includegraphics[width=67mm]{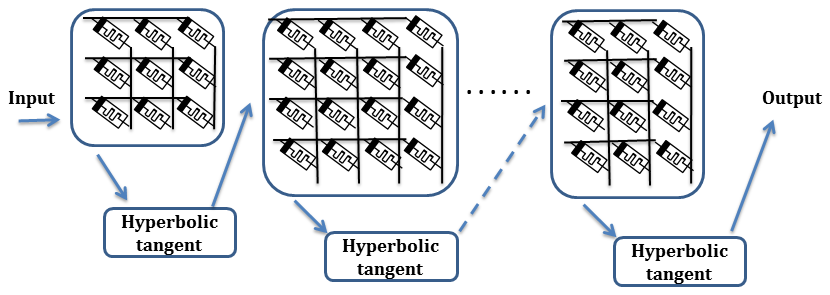}}
    \label{LSTM}
    \subfigure[]{ 
    \centering
   \includegraphics[width=70mm]{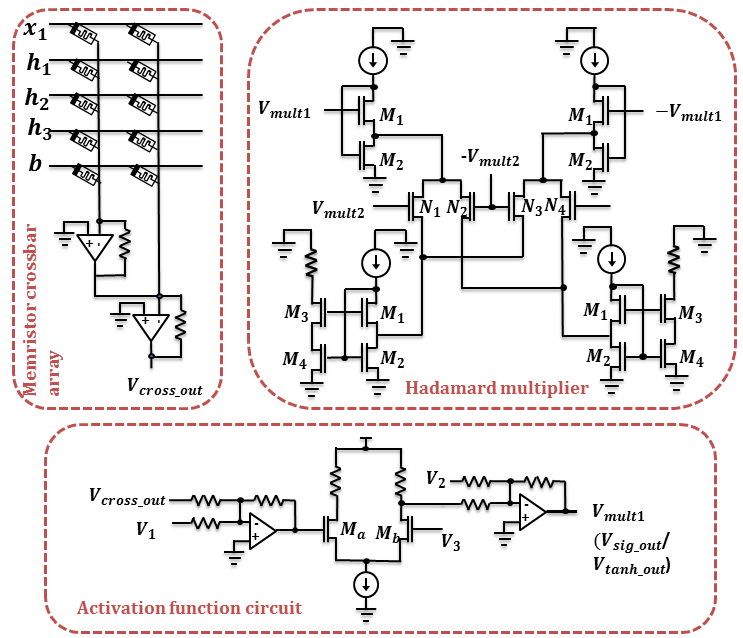}}
     \label{HTM}
    \subfigure[]
   { \includegraphics[width=65mm]{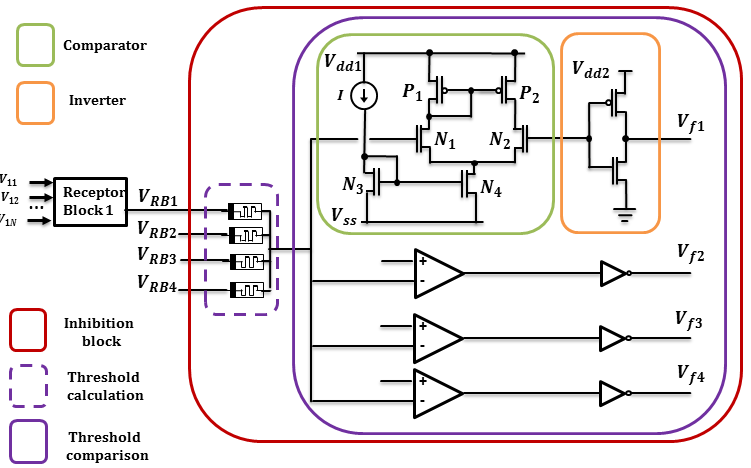} }
     \label{DNN}
     \caption{Circuit designs of a) DNN b) LSTM gate and c) Modified HTM }
     \label{3T}
\end{figure}

This paper analyses and compares the ability of hybrid CMOS-memristive neural architectures such as perceptron, ANN, DNN, LSTM and modified HTM to perform wafer classification. Fig. 1 illustrates the circuit blocks of compared neuromorphic memristive architectures. The weights of these architectures are implemented using memristive devices, while neurons and other computational components are using MOSFETs in TSMC CMOS process technology.

Single layer perceptrons are linear binary classifier and inspired from the biological neuron proposed in \cite{rosenblatt1958perceptron}. In this work, we used a single layer perceptron that has 151 input values corresponding to the size of the database patterns and 1 output neuron with hyperbolic tangent ($tanh$) activation function. The output of the network is $y_i=tanh(\sum x_{ij}*w_{ij})$, where $y_i$ is an output neuron, $x_{ij}$ are inputs to the perceptron and $w_{ij}$ are the weights in the network.
Three layer ANN, used in this work, consists of input, output and hidden layers with 151 inputs, 300 hidden layer neurons and single output neuron showing the binary classes of good and bad wafers. While DNN shown in Fig. 1 (a) has larger number of layers. In this work, we investigated the performance of 5 layer DNN with 151-300-50-100-1 nodes in each layer. The $tanh$ activation function was used in both ANN and DNN. 
\par
While ANN and DNN rely on simplistic multilayered information processing architectures, LSTM and HTM are more complex and emulate more information processing process in a human brain incorporating the concepts of gated memories and contextual processing.
LSTM is a neural architecture that uses the gated computation approach, and LSTM output depends on the previous state. Fig. 1 (b) illustrates the memristive hardware implementation of gated LSTM proposed in \cite{smagulova2018memristor}, \cite{smagulova2018design}.
In this work, classification using LSTM was performed using two layer network topology: LSTM layer with LSTM units unrolled for 151 time-steps and ANN layer with linear activation function. 


Hierarchical temporal memory is a neural architecture that mimics computational process in cerebral neocortex \cite{hawkinsintelligence}. HTM consists two main parts: (1) Spatial Pooler (SP) that performs encoding of input patterns and outputs Sparse Distributed Representation (SDR) useful for pattern recognition and image processing tasks, and (2) Temporal Memory (TM) that can be used for sequence learning and prediction. There are several hardware implementations of HTM. In this work, we explore analog hardware implementation of modified memristive HTM proposed in \cite{8023844} and represented in Fig. 1 (c).

\begin{table*}[hh]

\centering
\label{t2}
\caption{Comparison of LSTM performance with ANN, DNN and HTM for $40\times 1000$ iterations.}
\begin{tabular}{|l|c|c|c|c|}
\hline
\textbf{Method}                                                                                   & \textbf{On-chip area}                                                                & \textbf{Power consumption}                                                        & \textbf{\begin{tabular}[c]{@{}c@{}}Classification \\ Accuracy\end{tabular}} & \textbf{Comments}                                                                                                                                             \\ \hline
\begin{tabular}[c]{@{}l@{}}LSTM + ANN layer\\ (sequential: 4 hidden\\ units, 1 input, 152\\ time steps)\end{tabular}                                       & \begin{tabular}[c]{@{}c@{}}$ 257,503.20 \mu m^2 $\\ (with non-ideal\\ current sources)\end{tabular}                                                                                     &\begin{tabular}[c]{@{}c@{}}$ 255.8mW $\\ (maximum input values\\ were scaled down to 0.5) \end{tabular}                                                                                   & \begin{tabular}[c]{@{}c@{}}$ 98.51\%  $\\  \end{tabular}                                                                        & \begin{tabular}[c]{@{}c@{}} Learns significantly slower than single LSTM layer\\ with 1 time step. Exhibits increasing accuracy as\\ epoch size is increased. Gave accuracy of 97.26\% for\\ 25 epochs, 98.51\% for 40 epochs, and 98.86\%\\ for 55 epochs \end{tabular}                                                                                                                                                              \\ \hline
\begin{tabular}[c]{@{}l@{}}Single LSTM layer\\ (parallel: 1 hidden\\ unit, 152 inputs, 1\\ time step)\end{tabular}                                         & \begin{tabular}[c]{@{}c@{}}$ 115,967.4 \mu m^2 $\\ (with non-ideal\\ current sources)\end{tabular}                                                                                     & \begin{tabular}[c]{@{}c@{}}$ 312.4mW $\\ (maximum input values\\ were scaled down to 0.1) \end{tabular}                                                                                     & 96.09 \%                                                                            & \begin{tabular}[c]{@{}c@{}} Exploits windowing method: each time step is \\ considered as a feature. Can give up to 99.29\%\\ of accuracy when epoch size is increased to 100.  \end{tabular}                                                                                             \\ \hline
\begin{tabular}[c]{@{}l@{}}Perceptron\\ (tangent \\ activation f-on)\end{tabular}                 & \begin{tabular}[c]{@{}c@{}}$ 2,994.00 \mu m^2 $\\ (without buffer)\end{tabular}          & \begin{tabular}[c]{@{}c@{}}$80 mW$\\ (without buffer)\end{tabular}                & 90 \%                                                                       & \begin{tabular}[c]{@{}c@{}}Learns faster than ANN, however cannot \\ reach accuracy more than 93.7\% even with increased \\ number of iterations\end{tabular} \\ \hline
\begin{tabular}[c]{@{}l@{}}3 layer ANN\\ (300 neurons \\ in a hidden layer)\end{tabular}          & $ 4,839.90 \mu m^2$                                                                    & $ 1072.4 mW$                                                                      & 83\%                                                                        & \begin{tabular}[c]{@{}c@{}}Converges slower than LSTM for the same\\  learning rate and number of training iterations\end{tabular}                            \\ \hline
\begin{tabular}[c]{@{}l@{}}DNN\\ (5 layers,\\ 300-50-100 neurons\\ in hidden layers)\end{tabular} & $ 0.0121 mm^2$                                                                       & $ 2681.1 mW$                                                                      & 64\%                                                                        & \begin{tabular}[c]{@{}c@{}}The number of iterations should be\\  increased to achieve higher accuracy\end{tabular}                                            \\ \hline
Modified HTM                                                                                               & \begin{tabular}[c]{@{}c@{}}$0.096$ $mm^2$\\ (for sequential processing)\end{tabular} & \begin{tabular}[c]{@{}c@{}}$1756$ $mW$\\ (for sequential processing)\end{tabular} & 50\%                                                                        & \begin{tabular}[c]{@{}c@{}}Not effective for the small number of \\ input features and not able to converge\end{tabular}                                      \\ \hline
\end{tabular}
\end{table*}


    


\par

\section{Results and discussion}

The system level simulations have been performed performed in Python and MATLAB, while circuit level simulations are performed in SPICE for TSMC 180nm CMOS process. To investigate the performance of neuromorphic architectures, we used the wafer database that includes 151 patterns of inline process control measurements collected from different sensors during silicon wafer fabrication process\cite{olszewski2001generalized}. 
The wafers are divided into two classes: normal and abnormal wafers. Approximately 14\% of patters were used for training and the remaining 86\% were involved in a testing process. The database is characterized with imbalance of data between classes, particularly approximately 90\% of both training and testing wafer samples are normal wafers.   
\par


\begin{table}[]
\centering
\label{t1}
\caption{Comparison of software and hardware results by LSTM.}
\begin{tabular}{|c|c|c|c|}
\hline
\textbf{\begin{tabular}[c]{@{}c@{}}Wafer test\\ number\end{tabular}}                                                                                   & \textbf{\begin{tabular}[c]{@{}c@{}}Predicted value \\ (analog) (-mV)\end{tabular}}                                                                & \textbf{\begin{tabular}[c]{@{}c@{}}Predicted value \\ (software) (1e-3)\end{tabular}} & \textbf{Class}                                                                                                                                             \\ \hline
\begin{tabular}[c]{@{}l@{}}23 \end{tabular}	& -171.5 & -423.7 & -1 \\ \hline
\begin{tabular}[c]{@{}l@{}}47 \end{tabular}	& 523.2 & 473.0 & 1 \\ \hline
\begin{tabular}[c]{@{}l@{}}7 \end{tabular}	& -247.4 & -507.4 & -1 \\ \hline
\begin{tabular}[c]{@{}l@{}}3 \end{tabular}	& 470.7 & 511.6 & 1 \\ \hline
\begin{tabular}[c]{@{}l@{}}3838 \end{tabular}	& -485.5 & -418.3 & -1 \\ \hline
\begin{tabular}[c]{@{}l@{}}193 \end{tabular}	& 501.84 & 497.0 & 1 \\ \hline
\begin{tabular}[c]{@{}l@{}}6157 \end{tabular}	& -247.3 & -460.1 & -1 \\ \hline
\begin{tabular}[c]{@{}l@{}}411 \end{tabular}	& 531.9 & 489.8 & 1 \\ \hline
\begin{tabular}[c]{@{}l@{}}1534 \end{tabular}	& -437.2 & -456.2 & -1 \\ \hline
\begin{tabular}[c]{@{}l@{}}4507 \end{tabular}	& 255.6 & 493.6 & 1 \\ \hline

\end{tabular}
\end{table}

Table I shows the performance analysis of different  memristive neural architectures for wafer quality inspection. For all the architectures, performance analysis was performed for 40 iterations with 1000 training patterns in each with learning rate $\lambda = 0.001$. LSTM architecture consumes less power comparing to ANN and DNN and allows to achieve highest accuracy for the same number of iterations. LSTM can be considered as a best alternative for wafer quality inspection task.
In perceptron, ANN and DNN, the verification were performed using $tanh$ activation function, as we observed that linear and sigmoid activation functions did not ensure the convergence of the neural network weights for wafer database. Perceptron with a single neuron shows better accuracy for $40 \times 1000$ iterations than ANN and DNN, as the number of weights that should be trained is smaller. However, even for $4000 \times 1000$ training epochs, the maximum accuracy that can be achieved is approximately 94 \%. Also, the DNN accuracy is lower than ANN due to the large number of layers that cannot be trained with the small number of iterations. For ANN and DNN, the performance accuracy can be increased by increasing the number of training iterations.  
Modified HTM method proposed in \cite{8023844} is not able to converge for the wafer database, as the number of inputs are limited and the inputs are the data from various sensors. HTM is more effective for pattern recognition from the images where there is a correlation between the features, than for the data from various separate sensors. The on-chip area and power dissipation for the corresponding circuits were calculated for offline learning circuits and pretrained memristive weights based on the data from \cite{8023844} and \cite{olganew}. 

\begin{figure}[]
\centering
\includegraphics[width=0.5\textwidth]{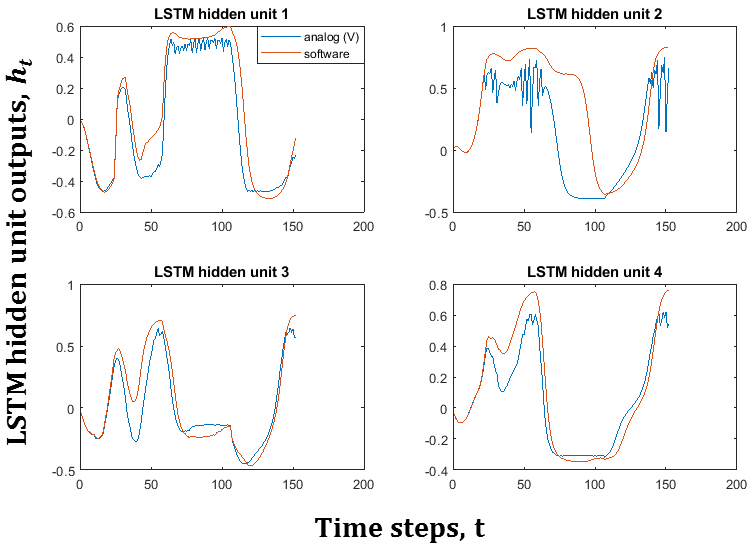}

\caption{LSTM cell outputs for analog and software implementations for test wafer 23}
\label{ht1}
\end{figure}

As LSTM demonstrates the best performance for wafer classification task, we investigate the system and circuit level performance of LSTM further. The simulation results for LSTM are shown in Table II and Fig. 2. Table II demonstrates the comparison of the LSTM results of system level and circuit level simulations for several exemplar wafers from the database.
Both of approaches exhibit positive or negative result depending on class; therefore, even though the circuit and software simulation results are different, the wafer classification can still be successfully performed.  Fig. 2 illustrates time dependent LSTM outputs for a single wafer, comparing software system level simulation and corresponding circuit level outputs in 4 hidden units.


\section{Conclusion}
In this work, wafer quality inspection and classification is performed using different neuromorphic memristive architectures.
LSTM outperforms the other architectures demonstrating the classification accuracy of 96-98\%. This can be explained by its gated structure that is capable of controlling the flow of information.  The same performance accuracy can be achieved by ANN and DNN, however more learning time, on-chip area and power consumption is required. A single perceptron has the smallest on-chip area and power consumption, however can not be trained to achieve high performance accuracy even with the large number of training iterations. Modified HTM for the given task has been found to be ineffective. Overall, neuromorphic memristive architectures can speed up the process of wafer quality inspection and can be integrated directly into sensors for measurements without sending data for the software processing and manual analysis. This can also result in reduced cost of wafer inspection.





\addtolength{\textheight}{-12cm} 

\bibliographystyle{IEEEtran} 
\bibliography{bib1}

\end{document}